\documentclass[11pt]{iopart}
\pdfoutput=1
\usepackage{cite,amssymb,amsfonts,graphicx,bm,dcolumn,mathrsfs}
\usepackage{color}
\usepackage{ulem}
\usepackage[utf8]{inputenc}
\usepackage[T1]{fontenc}
\usepackage{lmodern}

\begin{document}

\newcommand{\dif}{\rmd}
\newcommand{\ee}{\rme}
\newcommand{\ii}{\rmi}
\newcommand{\kB}{k_\mathrm{B}}
\newcommand{\sub}[1]{\mathrm{#1}}
\newcommand{\vect}[1]{\bi{#1}}

\title{Phase transitions in the unconstrained ensemble}
\author{Alessandro Campa$^1$, Lapo Casetti$^{2,3}$, Ivan Latella$^4$ and Stefano Ruffo$^{5,6}$}
\address{$^1$ National Center for Radiation Protection and Computational Physics, Istituto Superiore di Sanit\`{a}, Viale Regina
Elena 299, 00161 Roma, Italy}
\address{$^2$ Dipartimento di Fisica e Astronomia, Universit\`a di Firenze,\\and INFN, Sezione di Firenze,
via G.\ Sansone 1, 50019 Sesto Fiorentino (FI), Italy}
\address{$^3$ INAF-Osservatorio Astrofisico di Arcetri, Largo E.\ Fermi 5, 50125 Firenze, Italy}
\address{$^4$ Departament de F\'isica de la Mat\`eria Condensada, Universitat de Barcelona, \\ Mart\'i i Franqu\`es~1, 08028 Barcelona, Spain}
\address{$^5$ SISSA, via Bonomea 265 and INFN, Sezione di Trieste, 34136 Trieste, Italy}
\address{$^6$ Istituto dei Sistemi Complessi, Consiglio Nazionale delle Ricerche,
via Madonna del Piano 10, 50019 Sesto Fiorentino, Italy}
\ead{\mailto{alessandro.campa@iss.it}, \mailto{lapo.casetti@unifi.it}, 
\mailto{ilatella@ub.edu} and \mailto{ruffo@sissa.it}}

\begin{abstract}
The unconstrained ensemble describes completely open systems in which energy, volume and number of particles fluctuate. Here we show that not only equilibrium states can exist in this ensemble, but also that completely open systems can undergo first-order phase transitions. This is shown by studying a modified version of the Thirring model with attractive and repulsive interactions and with particles of finite size. The model exhibits first-order phase transitions in the unconstrained ensemble, at variance with the analogous model with point-like particles. While unconstrained and grand canonical ensembles are equivalent for this model, we found inequivalence between the unconstrained and isothermal-isobaric ensembles. By comparing the thermodynamic phase diagram in the unconstrained case with that obtained in the isothermal-isobaric ensemble, we show that phase transitions under completely open conditions for this model are different from those in which the number of particles is fixed, highlighting the inequivalence of ensembles.
\end{abstract}



\section{Introduction}
\label{intro}

Systems with long-range interactions are those where the interaction between two particles decays,
at large distance $r$, as $1/r^\alpha$, with $\alpha$ smaller than the space dimension $d$. These systems are intrinsically nonadditive; the interaction energy between two parts of the system cannot be neglected with respect to the energies of the two parts, independently of the size of the parts and of the system itself, and therefore also in the thermodynamic limit. 
This has important consequences on the equilibrium and nonequilibrium
properties \cite{Campa_2014,Campa_2009,Levin_2014,Bouchet_2010}. Far from just being a subject of academic interest, the study of
these systems rests its importance on the realization that there are numerous examples of long-range-interacting systems in nature: self-gravitating
systems \cite{Antonov_1962,Lynden-Bell_1968,Thirring_1970,Padmanabhan_1990,Lynden-Bell_1999,Chavanis_2002,deVega_2002_a,deVega_2002_b,
Chavanis_2006}, plasmas \cite{Nicholson_1992,Kiessling_2003}, two-dimensional fluid dynamics \cite{Onsager_1949,Miller_1990,
Robert_1991,Chavanis_2002_b,Venaille_2012} and spin systems \cite{Mori_2010,Mori_2011}.

Perhaps the most striking feature of nonadditive systems in equilibrium is the possible presence of ensemble inequivalence \cite{Thirring_1970,
Ellis_2000,Barre_2001,Bouchet_2005}, that physically means that the equilibrium properties of the system depends on which thermodynamic
parameters are taken as control variables, contrary to what happens for additive systems. Ensemble inequivalence is related to anomalous
concavities of the thermodynamic potentials associated with the various ensembles, this fact giving rise to negative response
functions \cite{Bouchet_2005,Chomaz_2002,entropy_2018}, most notably the possibility of a negative specific heat in the microcanonical ensemble.
Concerning the nonequilibrium behaviour, long-range isolated systems may remain trapped in nonequilibrium quasistationary states, with a lifetime diverging with the number of particles $N$, although for large but finite $N$ they eventually evolve towards
equilibrium \cite{Campa_2014,Campa_2009,Levin_2014,Yamaguchi_2004,Pereira_2012,Giachetti_2019}. 

There is another important feature that pertains exclusively to the equilibrium properties of nonadditive systems, related
to the violation of the Gibbs-Duhem equation \cite{Latella_2013,Latella_2015}. It is the possibility to have stable thermodynamic
states in completely open situations, that occur when none of the mechanical variables $E$ (total energy of the system), $V$ (volume
containing the systems) and $N$ (total number of particles) is a control parameter; in other words, when all these variables,
that can be collectively denoted as constraint variables, are not
held fixed but are allowed to fluctuate. We recall that when all the constraint variables are control parameters, the system is studied
in the microcanonical ensemble which can therefore be denoted as a completely constrained ensemble.  The case in which $E$, $V$
and $N$ fluctuate is then the opposite situation, and the corresponding ensemble can be denoted as the unconstrained ensemble.
However, it turns out that this ensemble cannot be defined for additive systems. Things go as follows. When only $E$ fluctuates, we
are in the canonical case in which the control parameters are $(T,V,N)$ (where $T$ is the temperature); if $E$ and $N$ fluctuate, we treat the system in the grand canonical ensemble, with control parameters $(T,V,\mu)$ (where $\mu$ is the chemical potential); if $E$ and $V$ fluctuate, we are studying the isothermal-isobaric ensemble, with control parameters $(T,P,N)$ (where $P$ is the pressure). One concludes that in the unconstrained ensemble the control parameters are $(T,P,\mu)$. But this cannot be realized in additive systems, since for
them $\mu$ is a function of $T$ and $P$, and therefore it cannot be used as an independent control parameter together with
$T$ and $P$. This is related to the fact
that for additive systems the thermodynamic potentials are linear homogeneous functions of the extensive variables, in particular of $V$
and $N$ \cite{Latella_2015}. This is also the origin of the Gibbs-Duhem relation.

On the contrary, the simple linear relation between the thermodynamic potentials and the extensive variables does not hold for
nonadditive systems, $\mu$ can be independent from $T$ and $P$ (although not in all cases \cite{Latella_2017}), and the possibility
to define the unconstrained ensemble arises. The theoretical framework introduced by Hill for small systems (which are nonadditive also
in the case of short-range interactions) \cite{Hill_2013,Hill_1998,Hill_2001,Hill_2001b} can be extended to long-range macroscopic
systems both in a purely thermodynamic approach \cite{Latella_2015}, and in a statistical mechanics treatment \cite{entropy_2018,
Latella_2017}, some details of which are given in Section \ref{uncon_ens}, where we introduce the unconstrained partition function
and the associated thermodynamic potential $\mathscr{E}$, the replica energy.

In \cite{Latella_2017} we have studied, in the framework of the unconstrained ensemble, a modified Thirring model. The model introduced
by Thirring \cite{Thirring_1970} is a simple model that reproduces some of the properties of self-gravitating systems~\cite{Campa_2016}. In
Section~\ref{themodel}, we recall the model and the modification considered in \cite{Latella_2017}. In the latter work we have shown
the inequivalence between the unconstrained and the canonical ensembles for this model. In particular, we have found that, while the model presents phase
transitions in the canonical ensemble, it does not have any transition in the unconstrained ensemble.

Stimulated by that result, the main purpose of this work is the study of a system that shows phase transitions in the unconstrained
ensemble. To the best of our knowledge, phase transitions in completely open nonadditive macroscopic systems have not been found so
far. To this end we have further extended the Thirring model, by considering particles of finite size. We also compare the thermodynamic
phase diagram in the unconstrained ensemble with that resulting from the study of the system in the two ensembles where two of the
constraint variables fluctuate, i.e., the isothermal-isobaric ensemble, where the control parameters are $(T,P,N)$, and the grand canonical
ensemble, where the control parameters are $(T,V,\mu)$. We have found that the unconstrained ensemble is inequivalent to the
isothermal-isobaric ensemble, while it is equivalent to the grand canonical ensemble.

The paper is organized as follows. In Section \ref{themodel} we describe the modified Thirring model, showing how a method introduced
by Thirring \cite{Thirring_1970} and here adapted for finite-size particles allows us to compute easily the configurational part of the partition functions. In Section \ref{uncon_ens}
we evaluate the unconstrained partition function and its associated thermodynamic potential, the replica energy. In Section \ref{isob_ens}
we compute the isothermal-isobaric partition function and the associated thermodynamic potential, the Gibbs free energy. In Section
\ref{grandcan_ens} we consider the grand partition function and the associated thermodynamic potential, the grand potential. The evaluations
in these last three sections are the basis for the determination of the thermodynamic phase diagrams in the respective ensembles. Before
determining the diagrams in Section \ref{phase_diag}, in Section \ref{landau_exp} we find their critical points, employing Landau's
theory of phase transitions. Finally, in Section \ref{discuss} we discuss the results and draw the conclusions. 

\section{The modified Thirring model with finite-size particles}
\label{themodel}

The modified Thirring model that we consider is a system of $N$ particles of mass $m$ and finite size characterized by the volume $\sigma$ and which occupy a volume $V$. The Hamiltonian of the system is
\begin{equation}
\mathcal{H}=\sum_{i=1}^N\frac{|\vect{p}_i|^2}{2m}+\sum_{i> j}^N\phi(\vect{q}_i,\vect{q}_j),
\label{Hamiltonian}
\end{equation}
$\vect{p}_i$ and $\vect{q}_i$ being the momentum and position of particle $i$, respectively, while $\phi(\vect{q}_i,\vect{q}_j)$
is the interaction potential given by
\begin{equation}
\phi(\vect{q}_i,\vect{q}_j) = 
-2\nu\left[ \theta_{V_0}(\vect{q}_i) \theta_{V_0}(\vect{q}_j)
+b\theta_{V_1}(\vect{q}_i) \theta_{V_1}(\vect{q}_j) \right],
\label{interaction_potential}
\end{equation}
where $\nu>0$ and $b$ are constants, and we use the functions $\theta_{V_k}(\vect{q}_i )$ defined by $\theta_{V_k}(\vect{q}_i ) =1$
if $\vect{q}_i\in V_k$  and $\theta_{V_k}(\vect{q}_i ) =0$ otherwise, with $k=0,1$. Here $V_0$ is the volume of an internal region of the
system ($V_0< V$) and $V_1$ is the complementary volume in such a way that $V= V_0 +V_1$. The position $\vect{q}_i$ of the $i$-th particle
is that of the center of its volume $\sigma$. The coupling constant $\nu$, coupling strength
$b$, core volume $V_0$ and particle volume $\sigma$ are all parameters of the model that do not depend on the state of the system.
In the original Thirring model \cite{Thirring_1970}, the parameters $b$ and $\sigma$ vanish, while in the modified model used
in Ref. \cite{Latella_2017} the parameter $b$ was introduced, but the particles where still point-like (i.e., $\sigma=0$).
\footnote{ In
the following we will consider the excluded volume of $N$ particles equal to $N\sigma$. If the particles
are thought of as hard spheres, $\sigma$ is equal to $4$ times the actual volume of the particles (see Section~\ref{local_eos}).}

In view of (\ref{interaction_potential}), the potential energy of the system is given by
\begin{equation}
\hat{W}(N_0,N_1)\equiv \sum_{i> j}^N\phi(\vect{q}_i,\vect{q}_j)
=-\nu\left( N_0^2+b N_1^2\right),
\label{potential_energy}
\end{equation}
where $N_0$ is the number of particles in $V_0$ and $N_1=N-N_0$ is the number of particles in $V_1$ for a given configuration. 
The canonical partition function thus reads
\begin{equation}
Z(T,V,N)
=\int\frac{\dif^{3N}\vect{q}}{\lambda_T^{3N} N!} \,\ee^{-\beta\hat{W}(N_0,N_1)},
\label{canonical_partition_function} 
\end{equation}
where $\beta=1/T$ is the inverse temperature, $\lambda_T=h/\sqrt{2\pi m T}$ is the thermal wave length, $h$ is a constant and we
have taken units in which the Boltzmann constant is $\kB=1$. In order to introduce excluded volume effects due to the finite size
of the particles, we use Thirring's trick to compute the integral over coordinates in such a way that
\begin{equation}
\int\frac{\dif^{3N}\vect{q}}{N!} \rightarrow\sum_{N_0,N_1}\delta_{N,N_0+N_1}\frac{(V_0-N_0\sigma)^{N_0}}{N_0!}
\frac{(V_1-N_1\sigma)^{N_1}}{N_1!},
\label{Thirring_method}
\end{equation}
where the sum runs over all possible values of $N_0$ and $N_1$ and the Kronecker $\delta$ enforces the relation $N=N_0+N_1$.
The prescription (\ref{Thirring_method}) does not consider configurations in which the volume $\sigma$ of one or more particles is partly contained in $V_0$ and partly in $V_1$. The contribution of these configurations to the partition function can be
neglected in the thermodynamic limit. Moreover, in Section~\ref{local_eos} we discuss in detail that (\ref{Thirring_method}) leads to the van der Waals equation of state for the local pressure.
The canonical partition function (\ref{canonical_partition_function}) computed using (\ref{Thirring_method}) will be the starting point in the sections below, where we study the model in the
unconstrained, isothermal-isobaric and grand canonical ensembles.

\section{The unconstrained ensemble}
\label{uncon_ens}

As mentioned in the Introduction, the unconstrained ensemble describes completely open systems in which all the constraint
variables (energy, volume and number of particles) fluctuate. The control parameters characterizing the state of the system are the
temperature $T$, pressure $P$ and chemical potential $\mu$. In the unconstrained ensemble, the replica energy $\mathscr{E}$ is the
associated free energy that rules equilibrium configurations. As shown in \cite{Latella_2017}, the unconstrained partition function
can be written as
\begin{eqnarray}
\Upsilon(T,P,\mu)
&=&\int \dif V\sum_N Z(T,V,N)\nonumber\\*
&=&\int \dif V\sum_N\int\frac{\dif^{3N}\vect{q}}{\lambda_T^{3N} N!} \,\ee^{\beta\mu N-\beta PV-\beta\hat{W}(N_0,N_1)}.
\label{completely_open_partition_function} 
\end{eqnarray}
The associated free energy, the replica energy $\mathscr{E}$, is defined by
\begin{equation}
\label{def_repl_ener}
\mathscr{E}(T,P,\mu) = - T \ln \Upsilon(T,P,\mu).
\end{equation}
It can be shown \cite{Latella_2015,Latella_2017} that in terms of the thermodynamic variables the replica energy can be expressed by
\begin{equation}
\mathscr{E} = E -TS +PV -\mu N ,
\label{expr_repl}
\end{equation}
where $S$ is the entropy of the system.
The first three terms on the right hand side constitute the Gibbs free energy $G$. We know that for additive systems $G=\mu N$, that
explains why in that case the replica energy vanishes. On the other hand, for nonadditive systems $G$ is not equal to $\mu N$, since the
thermodynamic potentials are not linear homogeneous functions of the constraint variables, and thus $\mathscr{E}$ is different from zero.

Using equation (\ref{Thirring_method}) to compute the integral over the coordinates, the unconstrained partition function
can be expressed as
\begin{equation}
\Upsilon=\int \dif V\sum_{N_0,N_1}\ \ee^{-\beta\hat{\mathscr{E}}(V,N_0,N_1)},
\label{completely_open_partition_function_3}
\end{equation}
where
\begin{equation}
\fl \hat{\mathscr{E}}(V,N_0,N_1) = PV
+T\sum_kN_k\left[\ln\left(\frac{N_k\lambda_T^3}{V_k-N_k\sigma}\right)-1-\frac{\mu}{T}\right] + \hat{W}(N_0,N_1).
\label{replica_energy_hat}
\end{equation}
From now on we omit writing down explicitly the dependence of the replica energy and other functions on the control parameters
$T$, $P$ and $\mu$. From equation
(\ref{def_repl_ener}), the replica energy can be obtained in the saddle-point
approximation as
\begin{equation}
\mathscr{E}=\inf_{\{V,N_0,N_1\}} \hat{\mathscr{E}}(V,N_0,N_1).
\label{replica_energy_minimization}
\end{equation}
Hence, minimization with respect to $V$, $N_0$ and $N_1$ yields the three equations
\begin{eqnarray}
P&=&\frac{T\bar{N}_1}{\bar{V}-V_0-\bar{N}_1\sigma},\label{sp_1}\\
\mu&=&-2\nu \bar{N}_0 
+T\ln\left(\frac{\bar{N}_0 \lambda_T^3}{V_0-\bar{N}_0\sigma}\right) + \frac{T\bar{N}_0\sigma}{V_0-\bar{N}_0\sigma},\label{sp_2}\\
\mu&=& -2\nu b \bar{N}_1
+T\ln\left(\frac{\bar{N}_1 \lambda_T^3}{\bar{V}-V_0-\bar{N}_1\sigma}\right) + \frac{T\bar{N}_1 \sigma}{\bar{V}-V_0-\bar{N}_1\sigma}\label{sp_3},
\end{eqnarray}
where $\bar{V}$, $\bar{N}_0$ and $\bar{N}_1$ are the values of the volume and of the number of particles in each region that
minimize the replica energy. The states of the system in the unconstrained ensemble are obtained by solving these equations for $\bar{V}$,
$\bar{N}_0$ and $\bar{N}_1$ as functions of the control parameters $T$, $P$ and $\mu$.

To study the equilibrium states of the system in this ensemble, it is convenient to introduce dimensionless quantities, as done
in \cite{Latella_2017}. Thus, we define
\begin{equation}
v=\frac{V-V_0}{V_0},\qquad x_0=\frac{\nu N_0}{T},\qquad x_1=\frac{\nu N_1}{T},
\qquad x=x_0+x_1,
\label{v_x_0_x_1}
\end{equation}
that will be written as $\bar{v}$, $\bar{x}_0$, $\bar{x}_1$ and $\bar{x}$ when evaluated at $\bar{V}$, $\bar{N}_0$ and $\bar{N}_1$.
We also introduce the exclusion parameter $a$, reduced pressure $p$ and chemical potential $\xi$ given by
\begin{equation}
a = \frac{T\sigma}{\nu V_0}, \qquad
p = \frac{\nu V_0}{T^2}P,\qquad 
\xi = \frac{\mu_T-\mu}{T},\qquad 
\label{p_z}
\end{equation}
where
\begin{equation}
\mu_T=T\ln\left(\frac{T\lambda_T^3}{\nu V_0}\right),
\end{equation}
and take advantage that using in the unconstrained ensemble the control parameters $a$, $p$ and $\xi$ is equivalent to using
$T$, $P$ and $\mu$. In \cite{Latella_2017}, the fugacity $z=\ee^{-\xi}$ was
used to control the states of the system instead of $\xi$.
We then write the dimensionless function $\hat{\varphi}_u=\nu\hat{\mathscr{E}}/T^2$ using the dimensionless quantities as
\begin{eqnarray}
\hat{\varphi}_u(v,x_0,x_1) &=&  
x_0\left[\ln\left(\frac{x_0 }{ 1 - a x_0 }\right)-1\right]
+ x_1\left[\ln\left(\frac{x_1}{v  -a x_1  } \right) -1\right] \nonumber\\
&& + p(v +1) + (x_0 + x_1)\xi - x_0^2 - b x_1^2.
\end{eqnarray}
In terms of this function, the variational problem (\ref{replica_energy_minimization}) becomes
\begin{equation}
\varphi_u=\inf_{\{v,x_0,x_1\}} \hat{\varphi}_u(v,x_0,x_1),
\label{replica_energy_minimization_2}
\end{equation}
where $\varphi_u=\nu\mathscr{E}/T^2$ is the dimensionless reduced replica energy in the unconstrained ensemble. The function
$\hat{\varphi}_u$ will also be called reduced replica energy or simply replica energy when no confusion arises, just to give it a name.
With the set of dimensionless variables, the saddle-point equations (\ref{sp_1})-(\ref{sp_3}) can be put as
\begin{eqnarray}
\bar{v} &=& \frac{1+ap}{2b p} \left(\ln p  + a p +\xi\right),\label{eq_1}\\
\bar{x}_0 &=&
\frac{1}{2}\ln\left(\frac{\bar{x}_0 }{1-a\bar{x}_0} \right) + \frac{a\bar{x}_0}{2(1-a\bar{x}_0)} + \frac{\xi}{2}, \label{eq_2}\\
\bar{x}_1 &=& 
\frac{1}{2b}\left(\ln p + a p +\xi \right),\label{eq_3}
\end{eqnarray}
where we have used that the reduced pressure is
\begin{equation}
p= \frac{\bar{x}_1}{\bar{v}-a\bar{x}_1 }. 
\end{equation}
Notice that $a\bar{x}_0=\bar{N}_0\sigma/V_0<1$ and $a\bar{x}_1/v=\bar{N}_1\sigma/V_1<1$, since these quantities represent the packing fraction
of particles in the core region $V_0$ and in the external region $V_1$, respectively.

The Hessian matrix $H_u$ associated to $\hat{\varphi}_u$ at the stationary point $(\bar{v},\bar{x}_0,\bar{x}_1)$ reads
\begin{equation}
\fl H_u=
\left(  \begin{array}{ccc} 
\bar{x}_1/(\bar{v} - a \bar{x}_1 )^2 &  0  &  - \bar{v} / (\bar{v}  -a \bar{x}_1)^2   \\
0 & 1/[\bar{x}_0(1-a\bar{x}_0)^2] -2   &  0\\
- \bar{v} / (\bar{v}  -a \bar{x}_1)^2 & 0   &  \bar{v}^2/[\bar{x}_1(\bar{v} -a\bar{x}_1)^2] -2b
\end{array} \right).
\end{equation}
Taking into account that $\bar{v}$, $\bar{x}_0$ and $\bar{x}_1$ are always positive quantities because of physical restrictions, this
matrix is positive-definite if and only if
\begin{equation}
 b <0, \qquad \frac{1}{2\bar{x}_0(1-a\bar{x}_0)^2} -1 >0,
 \label{cond3}
\end{equation}
which constitute the equilibrium conditions for the states obtained by solving the system of equations (\ref{eq_1})-(\ref{eq_3}).
We note that the problem decouples in two parts:  the minimization with respect to $v$ and $x_1$ and the minimization with respect to
$x_0$ alone. Correspondingly, the first of the conditions (\ref{cond3}) refers to the first part, and the second condition to the second
part. The full minimization problem then reduces to the minimization with respect to $x_0$ when $\hat{\varphi}_u$ is evaluated at
$v=\bar{v}$ and $x_1=\bar{x}_1$ satisfying the saddle-point equations. This fact allows us to study only the terms of
$\hat{\varphi}_u(\bar{v},x_0,\bar{x}_1)$ which depend on $x_0$, that is
\begin{equation}
\widetilde{\varphi}_u(x_0) = 
x_0\left[\ln\left(\frac{x_0 }{ 1 - a x_0 }\right)+\xi -1\right] - x_0^2.
\label{res_replica_energy}
\end{equation}
We note that this function vanishes for $x_0=0$ and that it tends to $+\infty$ for $x_0 \to 1/a$ from the left.
The extrema of $\widetilde{\varphi}_u$ are given by equation (\ref{eq_2}), while the condition to be a minimum is given by the second inequality in
(\ref{cond3}), that can be cast in the form
\begin{equation}
f_u(\bar{x}_0)=- 2 a^2\bar{x}_0^3 + 4 a\bar{x}_0^2 - 2\bar{x}_0 + 1   > 0.
\label{f_U}
\end{equation}
The function $f_u(\bar{x}_0)$ has the same sign of the second derivative of $\widetilde{\varphi}_u(x_0)$, and it can have one, two
(one simple and one double) or three real roots depending on the value of $a$.
When $a>8/27$, it has only one real root, larger than $1/a$, therefore in the physical relevant range $0<x_0<1/a$ the second derivative
of $\widetilde{\varphi}_u(x_0)$ is always positive, so that it can have at most one minimum. On the other hand, for $0<a<8/27$ the
function $f_u(\bar{x}_0)$ has three real roots, two in the range $0<x_0<1/a$ and one for $x_0>1/a$. Depending on the value of $\xi$,
it is then possible for the function $\widetilde{\varphi}_u(x_0)$ to have two minima and one maximum in the range $0<x_0<1/a$. If by
varying $\xi$ the height of these two minima exchange the role of lower minimum, the system can pass from one of these minima to the
other by undergoing a first-order phase transition. We note that in the limit $a \to 0$, the second condition in (\ref{cond3})
reduces to $\bar{x}_0<1/2$, and equation (\ref{eq_2}) can have at most one solution
in the range $0<x_0<1/2$, so that there cannot be phase transitions \cite{Latella_2017}.

\begin{figure}
\centering
\includegraphics[scale=0.85]{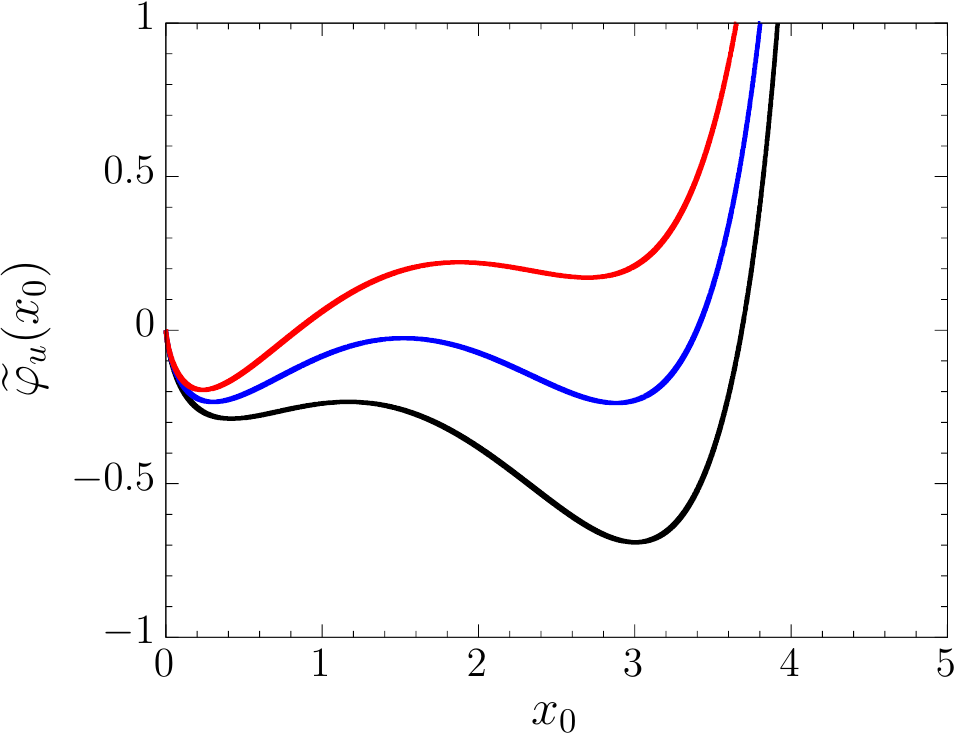}
\caption{Rescaled replica energy $\widetilde{\varphi}_u(x_0)$ as a function of $x_0$ in the uncosntrained ensemble, as given by equation
(\ref{res_replica_energy}), for $\xi= 1.5$ (black), $\xi= 1.65$ (blue) and $\xi= 1.8$ (red), taking $a=0.23$. The two relative minima
exchange their role as gobal minimum when $\xi$ is increased, showing that the system undergoes a first-order phase transition.}
\label{fig1}
\end{figure}

According to equations (\ref{eq_1}) and (\ref{eq_3}) and because $b<0$, the reduced pressure and chemical potential must be taken as
\begin{equation}
\ln p + a p < -\xi
\label{condition_pressure}
\end{equation}
to ensure that $\bar{v}$ and $\bar{x}_1$ are positive quantities, so that the above relation constitutes another equilibrium condition
in the unconstrained ensemble. 

As noted, the system can undergo first-order phase transitions that we now illustrate in figure~\ref{fig1} for some values of $\xi$. In
this figure, the two relative minima of $\widetilde{\varphi}_u(x_0)$ exchange their role as gobal minimum when $\xi$ is increased.

\subsection{Local equations of state}
\label{local_eos}

In order to investigate more closely the nature of excluded volume effects in the system, below we derive the local equations of state. As already explained, the system consists of two regions of volume $V_0$ and $V_1$ in which interactions are spatially constant in each of them. The density of particles in each region is thus spatially homogeneous, but, in general, the density in one region is different from the other. The local equations of state describe the thermodynamics in each of these regions, taking into account that global constraints apply to the whole system and therefore affect individual regions as well. As shown in the following, these equations can be obtained from the entropy by means of an usual thermodynamic approach. Thus, the first step is to obtain an expression for the entropy in the unconstrained ensemble.

We recall that the replica energy $\mathscr{E}$ is a function of the temperature, pressure and chemical potential. Variations of $\mathscr{E}$ and variations of $T$, $P$ and $\mu$ are related through the differential equation~\cite{Latella_2017}
\begin{equation}
\dif \mathscr{E} = -S\dif T + \bar{V} \dif P -\bar{N} \dif \mu, 
\end{equation}
which is the generalization of the Gibbs-Duhem equation for nonadditive systems. In our case $\bar{N}=\bar{N}_0+\bar{N}_1$. Thus, in the unconstrained ensemble, the entropy is given by
\begin{equation}
S=-\left( \frac{\partial \mathscr{E}}{\partial T} \right)_{P,\mu}
= \left. -\frac{\partial \hat{\mathscr{E}}}{\partial T}\right|_{\bar{V},\bar{N}_0,\bar{N}_1},
\label{entropy}
\end{equation}
where the expression on the right must be evaluated with $\bar{V}$, $\bar{N}_0$ and $\bar{N}_1$ satisfying the saddle-point equations (\ref{sp_1}), (\ref{sp_2}) and (\ref{sp_3}).
From equations (\ref{replica_energy_hat}) and (\ref{entropy}), we get
\begin{equation}
S = -\sum_k\bar{N}_k\left[\ln\left(\frac{\bar{N}_k\lambda_T^3}{\bar{V}_k-\bar{N}_k\sigma}\right)- \frac{5}{2 } \right],
\label{entropy_1}
\end{equation}
with $k=0,1$ and where we have used that $\lambda_T=h/\sqrt{2\pi m T}$. Here we also use the notation $\bar{V}_0=V_0$.

Let us now introduce the local specific volume $ v_k =\bar{V}_k/\bar{N}_k$ and the local internal energy per particle $u_k=3T/2$. The latter only accounts for kinetic degrees of freedom, since the long-range interactions in the system are seen by single particles as an external field. In terms of these local quantities, the entropy (\ref{entropy_1}) can be rewritten as
\begin{equation}
S = \sum_k\bar{N}_k s_k(u_k,v_k),
\label{entropy_2}
\end{equation}
where the local entropy per particle takes the form
\begin{equation}
s_k (u_k,v_k)= \ln\left[ c ( v_k -\sigma) u_k^{3/2}  \right] + \frac{5}{2 } ,
\label{local_entropy}
\end{equation}
with $c=(4\pi m /3h^2)^{3/2}$ a constant. The local equations of states are then given by the usual thermodynamic relations
\begin{equation}
\frac{1}{T_k}=\left( \frac{\partial s_k}{ \partial u_k} \right)_{v_k},\qquad 
\frac{p_k}{T_k}=\left( \frac{\partial s_k}{ \partial v_k} \right)_{u_k}.
\label{eq_T_k_p_k}
\end{equation}
The first of these equations in our case just states that the system is isothermal, since local temperatures $T_k$ correspond to that of the global thermostat, $T_k=T$. The second of the equations (\ref{eq_T_k_p_k}) specifies the local pressure $p_k$ and can be written as
\begin{equation}
 p_k(1-n_k\sigma)=n_kT,
 \label{van_der_Waals}
\end{equation}
where $n_k=1/v_k=\bar{N}_k/\bar{V}_k$ is the local number density. In view of the saddle-point equation (\ref{sp_1}), one readily notices that $P=p_1$, since $P$ is the pressure imposed at the boundary of the system and $p_1$ is the local pressure in its outer region. Moreover, using local variables, the remaining saddle-point equations can be put as
\begin{equation}
\mu= \psi_k + \mu_k,
\label{chemical_potential}
\end{equation}
where $k=0$ corresponds to equation~(\ref{sp_2}) and $k=1$ to equation~(\ref{sp_3}). Here we have introduced the local chemical potential
\begin{equation}
\mu_k= T\ln\left(\frac{p_k \lambda_T^3}{T}\right) + p_k \sigma
\end{equation}
and the potential energy $\psi_k=-2\nu \bar{N}_k q_k $, where $q_0=1$ and $q_1=b$. We note that $\psi_k$ can be regarded as the potential energy in region $k$ because the total potential energy can be computed as $W=\frac{1}{2}\sum_k \bar{N}_k \psi_k$. Notice that the chemical potential is spatially constant, since it takes the same value in the two regions of the system, ensuring that at equilibrium there is no net flux of particles between these regions and between the system and the reservoir.

Furthermore, it turns out that the local equation of state (\ref{van_der_Waals}) is the van der Waals equation of state with the parameter describing the van der Waals force equal to zero. This equation arises here as a consequence of the prescription (\ref{Thirring_method}), which has been employed to compute the partition function in the saddle-point approximation. Accordingly, the volume $\sigma$ introduced by this prescription has to be identified with $4\hat{\sigma}$ if the particles are hard spheres of volume $\hat{\sigma}$ (the excluded volume shared by two particles is a sphere of radius $2\hat{\sigma}$).
We highlight that the equation of state (\ref{van_der_Waals}) and the associated chemical potential (\ref{chemical_potential}) are the same used in reference~\cite{Aronson_1972}, in the context of long-range interacting systems, to describe a gas of self-gravitating hard spheres. In that case, the gravitational potential energy $\psi(r)$ is used instead of $\psi_k$ in a continuum description. In~\cite{Aronson_1972} (see also~\cite{Padmanabhan_1990}), the van der Waals equation of state is introduced in a thermodynamic framework at the local level, and global thermodynamic quantities are built on the basis of the local description.

We finally remark that the procedure here described to obtain the local equations of state can be implemented in other ensembles as well, but the corresponding entropy associated to the considered ensemble must be taken as the starting point. Along these lines, equations (\ref{eq_T_k_p_k}), in particular (\ref{van_der_Waals}), will have the same form, but quantities such as $T$ and $n_k$ will correspond to the actual physical conditions imposed on the system.  

\section{The isothermal-isobaric ensemble}
\label{isob_ens}

In this ensemble the control parameters defining the state of the system are $T$, $P$ and $N$. The equilibrium
configurations follow from the Gibbs free energy $G(T,P,N)=-T\ln \Delta(T,P,N)$, where the isothermal-isobaric partition function
$\Delta(T,P,N)$ can be obtained from the canonical partition function (\ref{canonical_partition_function}) as
\begin{eqnarray}
\Delta(T,P,N)
&=&\int \dif V\,\ee^{-\beta PV} Z(T,V,N)\nonumber\\
&=&\int \dif V\int\frac{\dif^{3N}\vect{q}}{\lambda_T^{3N} N!} \,\ee^{-\beta PV-\beta\hat{W}(N_0,N_1)}.
\label{gibbs_partition_function} 
\end{eqnarray}
Using Thirring's trick (\ref{Thirring_method}) again together with the fact that $N$ is fixed, we get
\begin{equation}
\Delta
=\int \dif V  \sum_{N_0}\ 
\ee^{- \beta\hat{G}(V,N_0)},
\end{equation}
where
\begin{equation}
\hat{G}(V,N_0)= PV
+T\sum_kN_k \left[\ln\left(\frac{N_k \lambda_T^3}{V_k-N_k \sigma}\right)-1\right] + \hat{W}(N_0,N_1)
\label{hat_G} 
\end{equation}
with $N_1=N-N_0$. As before, we omit writing down explicitly the dependence on the control parameters $T$, $P$ and $N$. The Gibbs
free energy is then given by
\begin{equation}
G=\inf_{\{V,N_0\}} \hat{G}(V,N_0),
\label{gibbs_free_energy_minimization}
\end{equation}
and the saddle-point equations take the form
\begin{eqnarray}
\bar{V}&=& V_0+\frac{(T+P\sigma)(N-\bar{N}_0)}{P},\label{sp_12}\\
\bar{N}_0 
&=& b\left(N-\bar{N}_0\right)
+\frac{T}{2\nu}\ln\left[\frac{T\bar{N}_0}{P(V_0-\bar{N}_0\sigma)}  \right] 
+ \frac{T\bar{N}_0\sigma }{2\nu(V_0-\bar{N}_0\sigma)}- \frac{P\sigma}{2\nu}\label{sp_22}
\end{eqnarray}
where now $\bar{V}$ and $\bar{N}_0$ are the volume and number of particles in region $V_0$ that solve the variational problem in the
isothermal-isobaric ensemble, and which are functions of the control parameters $T$, $P$ and $N$.

We introduce the reduced variables
\begin{equation}
v=\frac{V-V_0}{V_0},\qquad x_0=\frac{\nu N_0}{T},\qquad x=\frac{\nu N}{T},
\label{reduced_variables_gibbs}
\end{equation}
where $v$ and $x_0$ will be written as $\bar{v}$ and $\bar{x}_0$ when evaluated at $\bar{V}$ and $\bar{N}_0$, respectively. As before,
in this ensemble we introduce the exclusion parameter $a$ and reduced pressure $p$ and chemical potential $\xi$ given by
\begin{equation}
a = \frac{T\sigma}{\nu V_0}, \qquad
p = \frac{\nu V_0}{T^2}P,\qquad 
\xi = \frac{\mu_T-\mu}{T},\qquad 
\label{p_z_gibbs}
\end{equation}
with 
\begin{equation}
\mu_T=T\ln\left(\frac{T\lambda_T^3}{\nu V_0}\right),
\end{equation}
and take $a$, $p$ and $x$ as the control parameters which is equivalent to consider $T$, $P$ and $N$.
The chemical potential in the isothermal-isobaric ensemble is given by
\begin{equation}
\mu=\left(\frac{\partial G}{\partial N}\right)_{T,P}=\left.\frac{\partial \hat{G}}{\partial N}\right|_{\bar{V},\bar{N}_0}, 
\end{equation}
where the expression on the right must be evaluated at $\bar{V}$ and $\bar{N}_0$ satisfying the saddle-point equations. This leads to
\begin{equation}
\mu=-2b\nu(N-\bar{N}_0)+P\sigma+T\ln\left(\frac{P\lambda_T^3}{T}\right),
\end{equation}
which can be rewritten as
\begin{equation}
\xi= 2b(x-\bar{x}_0)-ap-\ln p.
\label{xi_gibbs}
\end{equation}
Analogously to the previous section, we introduce the dimensionless quantity $\hat{\varphi}_i=\nu\hat{G}/T^2$, that in terms of
the dimensionless variables is given by
\begin{eqnarray}
\hat{\varphi}_i(v,x_0) &=&  
x_0 \ln\left(\frac{ x_0 }{1-a x_0 } \right)  
- (x- x_0) \left[ \ln\left( \frac{v }{ x- x_0 } - a \right) - \frac{ p v }{ x- x_0} \right] \nonumber\\
&&  
- x_0^2 - b(x-x_0)^2 
+ p + x\left(\frac{\mu_T}{T} -1\right).
\label{varphi_I}
\end{eqnarray}
The corresponding variational problem is
\begin{equation}
\varphi_i=\inf_{\{v,x_0\}} \hat{\varphi}_i(v,x_0),
\label{gibbs_free_energy_minimization_2}
\end{equation}
where $\varphi_i=\nu G/T^2$ is the dimensionless reduced Gibbs free energy in the isothermal-isobaric ensemble. The function
$\hat{\varphi}_i$ will also be called reduced Gibbs free energy or simply Gibbs free energy when no confusion arises.
With the dimensionless variables, the saddle-point equations (\ref{sp_12}) and (\ref{sp_22}) become
\begin{eqnarray}
\bar{v}&=& \frac{(1+ap)}{p}(x-\bar{x}_0),\label{sp_12_reduced}\\
\bar{x}_0 
&=& 
\frac{1}{2}\ln\left(\frac{\bar{x}_0}{1-a\bar{x}_0}  \right)
+ \frac{a\bar{x}_0 }{2(1-a\bar{x}_0)}
+b\left(x-\bar{x}_0\right) - \frac{1}{2}\left( ap +\ln p\right). \label{sp_22_reduced}
\end{eqnarray}
Notice that the last two terms in equation (\ref{sp_22_reduced}) are equal to $\xi/2$, with $\xi$ being given by equation
(\ref{xi_gibbs}). If $\xi$ was a fixed quantity here, equations (\ref{eq_2}) and (\ref{sp_22_reduced}) would coincide; the fact that
these equations are different is a signature of a nonequivalence between the unconstrained and the isothermal-isobaric ensembles.
Moreover, the Hessian matrix $H_i$ associated to $\hat{\varphi}_i$ at the stationary point $(\bar{v},\bar{x}_0)$ takes the form
\begin{equation}
\fl H_i=
\left(  \begin{array}{cc} 
p^2/(x-\bar{x}_0)   &   p(1+ap)/(x-\bar{x}_0) \\
 p(1+ap)/(x-\bar{x}_0)   &  1/[\bar{x}_0(1-a\bar{x}_0)^2]-2(1+b) + (1+ap)^2/(x-\bar{x}_0)
\end{array} \right).
\end{equation}
Since $x-\bar{x}_0>0$ by definition, this matrix is positive-definite if and only if
\begin{equation}
\frac{1}{\bar{x}_0(1-a\bar{x}_0)^2} -2(1+b)>0,
\end{equation}
which constitutes the equilibrium condition in the isothermal-isobaric ensemble. The above condition can be rewritten as
\begin{equation}
f_i(\bar{x}_0)=-2 a^2 (1+b)\bar{x}_0^3 + 4a(1+b)\bar{x}_0^2 - 2(1+b)\bar{x}_0 +1 >0, 
\label{f_I}
\end{equation}
whose roots determine the permitted ranges for $\bar{x}_0$ in equilibrium configurations. In particular, for $0<a<8(1+b)/27$ with
$b>-1$, $f_i(\bar{x}_0)$ has three different real roots and therefore, the equilibrium condition leads to two disconnected regions for
possible values of $\bar{x}_0$. 
It is possible to reduce the dimensionality of the problem by evaluating $\hat{\varphi}_i(v,x_0)$ with either $v$ or $x_0$ at the
minimum. Although $\bar{v}$ and $\bar{x}_0$ are coupled through equations (\ref{sp_12_reduced}) and (\ref{sp_22_reduced}), this can be
achieved by noting that the ratio $v/(x-x_0)$ remains constant at equilibrium, as can be seen in equation (\ref{sp_12_reduced}).
According to this, using equation (\ref{sp_12_reduced}) in (\ref{varphi_I}) leads to
\begin{equation}
\widetilde{\varphi}_i(x_0) =  
x_0 \left[ \ln\left(\frac{ x_0 }{1-a x_0 } \right) - \ln p - ap - 1\right]
- x_0^2 - b(x-x_0)^2,
\label{varphi_I_2}
\end{equation}
where, for simplicity, a term that does not depend on $x_0$ has been omitted on the right hand side. The condition of minimum of
$\widetilde{\varphi}_i(x_0)$ is again equation (\ref{f_I}). Making an analogous reasoning as in the previous section, we argue that
when $f_i(\bar{x}_0)$ has two roots in the range $0<\bar{x}_0<1/a$, i.e., for $0<a<8(1+b)/27$ with $b>-1$, first-order phase
transitions may occur in this ensemble.

\begin{figure}
\centering
\includegraphics[scale=0.85]{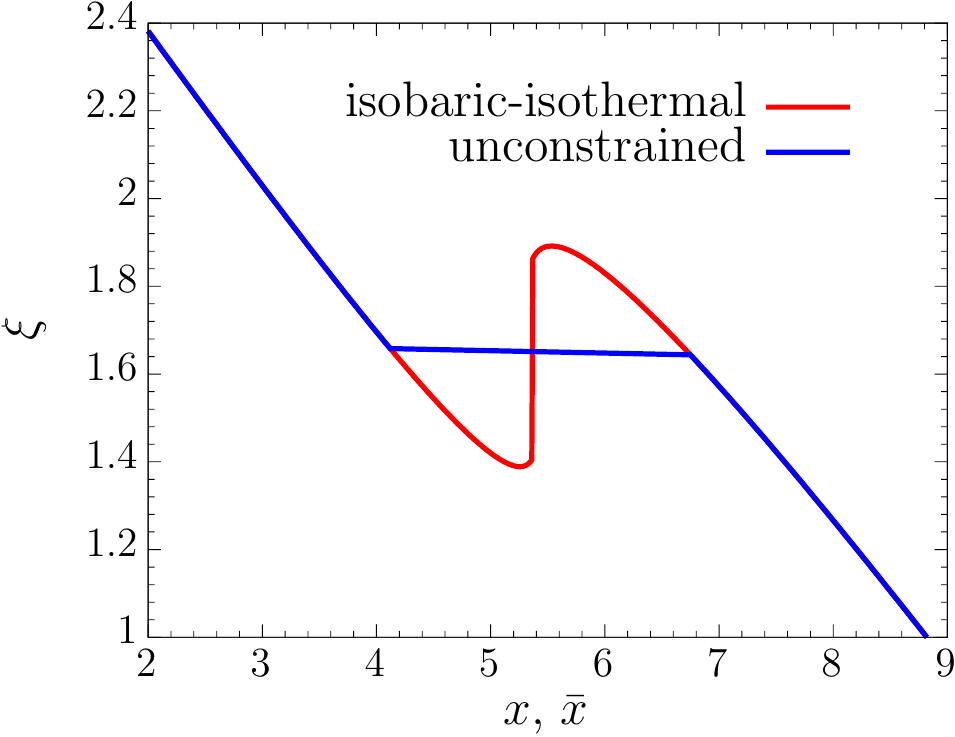}
\caption{Curves relating the total number of particles $x$ and $\bar{x}$ in the isothermal-isobaric and unconstrained ensembles,
respectively, to the corresponding chemical potential $\xi$. Here, the exclusion parameter is $a=0.23$, the pressure is $p=0.045$ and
the coupling is $b=-3/16$.}
\label{fig2}
\end{figure}

In order to draw a comparison between some observables in the unconstrained and isothermal-isobaric ensembles, in figure~\ref{fig2} we
show the curves relating the total number of particles to the chemical potential in the two ensembles. In the unconstrained case,
$\bar{x}$ is computed as a function of $\xi$, the latter being a control parameter in this ensemble. Instead, in the isothermal-isobaric
ensemble, $\xi$ is computed as a function of $x$ which is the control parameter in this case. A first-order transition is shown in both
cases, as well as a clear ensemble nonequivalence. In these curves, the exclusion parameter is $a=0.23$, the pressure is $p=0.045$ and
the coupling strength in the outer region of the system is $b=-3/16$. We note from the figure that in the isothermal-isobaric ensemble,
apart from the jump in $\xi$ at the first-order transition, there are small ranges in $x$, on both sides of the jump, where the
derivative of $\xi$ with respect to $x$ is positive. From the definition of the reduced  variable $\xi$ in (\ref{p_z_gibbs}), we see
that the sign of its derivative is the opposite of that of the chemical potential $\mu$; therefore in those small ranges the derivative
of $\mu$ with respect to $x$ is negative. It can be shown \cite{entropy_2018} that this derivative is positive definite in the
unconstrained ensemble, while it may be negative in the isothermal-isobaric ensemble. Thus, the equilibrium states of this model give
a concrete realization of that anomaly.

\section{The grand canonical ensemble}
\label{grandcan_ens}

In the grand canonical ensemble, the control parameters defining the state of the system are $T$, $V$ and $\mu$, and equilibrium
configurations follow from the grand potential $\Omega(T,V,\mu)=-T\ln \Xi(T,V,\mu)$. The grand canonical partition function
$\Xi(T,V,\mu)$ can be obtained from the canonical partition function (\ref{canonical_partition_function}) as
\begin{eqnarray}
\Xi(T,V,\mu)
&=&\sum_N \ee^{\beta \mu N} Z(T,V,N)\nonumber\\
&=&\sum_N \int\frac{\dif^{3N}\vect{q}}{\lambda_T^{3N} N!} \,\ee^{\beta \mu N-\beta\hat{W}(N_0,N_1)}.
\label{grand_partition_function} 
\end{eqnarray}
Using (\ref{Thirring_method}), this partition function can be written as
\begin{equation}
\Xi=\sum_{N_0,N_1}e^{-\beta\hat{\Omega}(N_0,N_1)},
\end{equation}
where
\begin{eqnarray}
\hat{\Omega}(N_0,N_1)&=&
T\sum_k N_k\left[\ln\left( \frac{N_k \lambda_T^3}{V_k - N_k \sigma}\right) -1 -\frac{\mu}{T}\right] 
+ \hat{W}(N_0,N_1)
\label{grand_potential}
\end{eqnarray}
and we have omitted writing down explicitly the dependence on $T$, $V$ and $\mu$. The grand potential is then given by the
variational problem
\begin{equation}
\Omega =\inf_{\{ N_0, N_1 \}} \hat{\Omega}(N_0,N_1),
\label{grand_potential_minimization}
\end{equation}
and the saddle-point equations in this case read as
\begin{eqnarray}
\mu&=&-2\nu \bar{N}_0 
+T\ln\left(\frac{\bar{N}_0 \lambda_T^3}{V_0-\bar{N}_0\sigma}\right) + \frac{T\bar{N}_0\sigma}{V_0-\bar{N}_0\sigma},\label{sp_1_grand}\\
\mu&=& -2\nu b \bar{N}_1
+T\ln\left(\frac{\bar{N}_1 \lambda_T^3}{ V-V_0-\bar{N}_1\sigma}\right) + \frac{T\bar{N}_1 \sigma}{ V-V_0-\bar{N}_1\sigma}\label{sp_2_grand},
\end{eqnarray}
where $\bar{N}_0$ and $\bar{N}_1$ are the number of particles in each region that minimize the grand potential. In the grand canonical
ensemble, the equilibrium states of the system are obtained by solving these equations for $\bar{N}_0$ and $\bar{N}_1$ as functions of
the control parameters $T$, $V$ and $\mu$.

Introducing dimensionless variables in the grand canonical ensemble, analogously defined as in equations (\ref{v_x_0_x_1})
and (\ref{p_z}), the reduced grand potential $\hat{\varphi}_g =\nu\hat{\Omega}/T^2$ takes the form
\begin{eqnarray}
\hat{\varphi}_g(x_0,x_1)&=&
x_0\left[\ln\left( \frac{ x_0 }{ 1 - a x_0 } \right) -1 \right]
+ x_1\left[\ln\left( \frac{ x_1 }{ v - a x_1  } \right) -1 \right] \nonumber\\
&+& ( x_0 + x_1 )\xi -  x_0^2 - b x_1^2,
\label{reduced_grand_potential}
\end{eqnarray}
and the variational problem (\ref{grand_potential_minimization}) becomes
\begin{equation}
\varphi_g =\inf_{\{ x_0, x_1 \}} \hat{\varphi}_g(x_0,x_1),
\end{equation}
where $\varphi_g =\nu \Omega/T^2$.
In terms of dimensionless variables, we take $a$, $v$ and $\xi$ as the control parameters, which is equivalent to taking $T$, $V$
and $\mu$ in the grand canonical ensemble. Furthermore, the Hessian matrix $H_g$ associated to $\hat{\varphi}_g$ at the stationary
point $(\bar{x}_0,\bar{x}_1)$ reads
\begin{equation}
H_g=
\left(  \begin{array}{ccc} 
1/[\bar{x}_0(1-a\bar{x}_0)^2] -2   &  0\\
0   &  v^2/[\bar{x}_1( v -a\bar{x}_1)^2] -2b
\end{array} \right).
\end{equation}
Thus $H_g$ is positive definite if and only if
\begin{equation}
\frac{v^2}{2\bar{x}_1( v -a\bar{x}_1)^2} -b > 0, \qquad 
\frac{1}{2\bar{x}_0(1-a\bar{x}_0)^2} -1 >0,
\label{equilibrium_grand}
\end{equation}
which constitute the equilibrium conditions in the grand canonical ensemble. In view of these conditions, we highlight that
$b\geq0$ can lead to equilibrium configurations in this ensemble. In the case $b<0$, the first of the conditions
(\ref{equilibrium_grand}) is always fulfilled and we are left with
\begin{equation}
b < 0, \qquad 
\frac{1}{2\bar{x}_0(1-a\bar{x}_0)^2} -1 >0.
\end{equation}
Thus, in this case, the equilibrium conditions in the grand canonical ensemble are exactly the same as in the unconstrained ensemble,
the latter being given by (\ref{cond3}). Below we argue that not only the equilibrium conditions are the same in this case, but also
that these two ensembles are equivalent when $b<0$. We recall that there are no equilibrium configurations in the unconstrained ensemble
when $b\geq0$.

To demonstrate such an equivalence, we only need to show that equilibrium states defined by $T$, $V$ and $\mu$ in the grand canonical
ensemble univocally correspond to equilibrium states characterized by $T$, $\bar{V}$ and $\mu$ in the unconstrained ensemble, with the
same pressure $P$ in the two cases.  We start by considering the grand canonical pressure which is given by
\begin{equation}
P =-\left(\frac{\partial \Omega}{\partial V}\right)_{T,\mu}=-\left.\frac{\partial \hat{\Omega}}{\partial V}\right|_{\bar{N}_0, \bar{N}_1}, 
\end{equation}
where the expression on the right must be evaluated at $\bar{N}_0$ and $\bar{N}_1$ satisfying the saddle-point
equations (\ref{sp_1_grand}) and (\ref{sp_2_grand}). Using (\ref{grand_potential}), we get
\begin{equation}
P = \frac{ T  \bar{N}_1 }{ V- V_0 - \bar{N}_1 \sigma }.
\label{P_grand}
\end{equation}
We observe that if the substitution $V\to \bar{V}$ is made in the above equation (\ref{P_grand}) and in equations (\ref{sp_1_grand})
and (\ref{sp_2_grand}), the saddle-point equations (\ref{sp_1}), (\ref{sp_2}) and (\ref{sp_3}) of the unconstrained ensemble are
obtained. Therefore, equilibrium states in the two ensembles are described by the same set of equations, establishing the correspondence
we wanted to show.

\section{Landau expansions and critical points}
\label{landau_exp}

The model exhibit first-order phase transitions in all the considered ensembles. In the next section we will describe the full phase
diagrams, and in this section we determine the critical points associated to the phase transitions. The critical points can be found
analytically with the use of Landau's theory by performing an expansion of the characteristic free energies in these
ensembles for $x_0$ around equilibrium configurations. Since we have found that the grand canonical ensemble is equivalent to the
unconstrained ensemble, in this section and in the next one we will make explicit evaluations only for the unconstrained and the
isothermal-isobaric ensembles. Thus, we consider a series of the form
\begin{equation}
\widetilde{\varphi}(m)=q_0+q_1m+q_2m^2+q_3m^3+q_4m^4+\mathcal{O}(m^5), 
\label{expansion}
\end{equation}
where the expansion parameter is $m=x_0-\bar{x}_0$ and $\widetilde{\varphi}(m)$ can be either the reduced replica energy
(\ref{res_replica_energy}) or the reduced Gibbs free energy (\ref{varphi_I_2}).

\subsection{Unconstrained ensemble}
\label{landau_exp_1}

To describe a critical point in this ensemble, we need to find the critical exclusion parameter, reduced pressure and chemical
potential at the end of a first-order transition line, which will be denoted as $a^u$, $p^u$ and $\xi^u$, respectively, together
with the critical value of the reduced number of particles in the core that will be denoted as $x_0^u$. The coefficients of the
expansion (\ref{expansion}) for the replica replica energy (\ref{res_replica_energy}) take the form
\begin{eqnarray}
q_0&=& \widetilde{\varphi}_u(\bar{x}_0), \label{q0_unconstrained}\\
q_1&=& \ln\left(\frac{\bar{x}_0 }{ 1 - a \bar{x}_0 }\right) 
+ \frac{a \bar{x}_0 }{ 1 - a \bar{x}_0 } +\xi - 2\bar{x}_0 ,\label{q1_unconstrained}\\
q_2&=& \frac{ f_u(\bar{x}_0) }{2\bar{x}_0(1-a\bar{x}_0)^2} ,\label{q2_unconstrained}\\
q_3&=& \frac{3 a \bar{x}_0 - 1 }{6\bar{x}_0^2(1-a\bar{x}_0)^3} ,\label{q3_unconstrained}\\
q_4&=& \frac{ 2a\bar{x}_0(3 a \bar{x}_0 - 2) +1 }{12\bar{x}_0^3(1-a\bar{x}_0)^4},\label{q4_unconstrained}
\end{eqnarray}
where $f_u(\bar{x}_0)$ is given by (\ref{f_U}). The critical conditions are obtained by setting $q_1=q_2=q_3=0$. From
(\ref{q3_unconstrained}), we see that $x_0^u=1/(3a^u)$, which replaced in (\ref{q2_unconstrained}) gives $a^u=8/27$ from the condition
$f_u\left(\frac{1}{3a^u}\right)=0$. Hence $x_0^u=9/8$. The critical value of the reduced chemical potential is readily obtained by
evaluating $q_1=0$ with these values for $x_0^u$ and $a^u$, yielding $\xi^u=7/4-\ln(27/16)\simeq 1.2267$. In addition, we have
$q_4=8/81$ under these conditions, ensuring that $\widetilde{\varphi}_u$ reaches a minimum there. Taking into account the definitions
of the reduced variables and the exclusion parameter, the value of the number of particles in the core $N_0^u$ and temperature $T^u$
at the critical point are given by
\begin{eqnarray}
N_0^u &= &\frac{V_0 }{3 \sigma} ,\\
T^u &= & \frac{8 \nu V_0 }{27  \sigma},
\end{eqnarray}
as well as the critical chemical potential difference
\begin{equation}
(\mu_T- \mu)^u = \frac{8 }{27 }\left( \frac{7}{4}-\ln \frac{27}{16}\right) \frac{ \nu V_0 }{ \sigma}.  
\end{equation}
We highlight that the critical conditions in the unconstrained ensemble do not depend on the particular value of the parameter $b$.
Notice also that these conditions are independent of the pressure. However, according to the equilibrium
condition (\ref{condition_pressure}), the pressure must obey $\ln p^u + a^u p^u < -\xi^u$ in order to maintain the system at the
critical point. This relation can be written as
\begin{equation}
\ln \left( \frac{8 p^u}{27} \right) + \frac{8 p^u}{27}  + \frac{7}{4} +\ln 2  <0,
\end{equation}
and can be numerically solved giving $p^u<0.2706$. We therefore have a line of critical points corresponding to each value of the
pressure fulfilling this condition. In the next section we show a phase diagram in the space of parameters $(p,a,\xi)$, in which
these critical points constitute a line parallel to the $p$ axis that terminates at $p=0.2706$. Finally, if the parameters of the
system are restored, the pressure $P^u$ at those critical points is defined by
\begin{equation}
 P^u< 0.0238 \frac{ \nu V_0 }{ \sigma^2} . 
\end{equation}

\subsection{Isothermal-isobaric ensemble}
\label{landau_exp_2}

The critical points in this ensemble are described by the critical exclusion parameter, reduced pressure and number of particles, which
will be denoted as $a^i$, $p^i$ and $x^i$, respectively. The associated critical number of particles in the core will be denoted
as $x_0^i$. In the isothermal-isobaric ensemble, the coefficients of the expansion (\ref{expansion}) for the Gibbs
free energy (\ref{varphi_I_2}) are given by
\begin{eqnarray}
q_0&=& \widetilde{\varphi}_i(\bar{x}_0), \label{q0_iso}\\
q_1&=& 
\ln\left( \frac{\bar{x}_0}{1-a\bar{x}_0} \right)
+ \frac{a\bar{x}_0 }{(1-a\bar{x}_0)}
+2b\left(x-\bar{x}_0\right) - \left( ap +\ln p\right) -2\bar{x}_0 ,\label{q1_iso}\\
q_2&=& \frac{ f_i(\bar{x}_0) }{2\bar{x}_0(1-a\bar{x}_0)^2} ,\label{q2_iso}\\
q_3&=& \frac{3 a \bar{x}_0 - 1 }{6\bar{x}_0^2(1-a\bar{x}_0)^3} ,\label{q3_iso}\\
q_4&=& \frac{ 2a\bar{x}_0(3 a \bar{x}_0 - 2) +1 }{12\bar{x}_0^3(1-a\bar{x}_0)^4},\label{q4_iso}
\end{eqnarray}
where $f_i(\bar{x}_0)$ is given by (\ref{f_I}). As before, the critical conditions are obtained by imposing $q_1=q_2=q_3=0$. From
$q_3=0$, we get $x_0^i=1/(3a^i)$, and using this quantity in the condition $q_2=0$ leads to $a^i=8(1+b)/27$ with $b>-1$. We therefore
have $x_0^i=9/(8+8b)$. Using these values of $x_0^i$ and $a^i$ in the equation $q_1=0$ for $b>-1$ yields the relation
\begin{equation}
2 b x^i ( p^i )=\frac{7}{4}+\frac{8(1+b)p^i}{27}+\ln\left[ \frac{16(1+b)p^i}{27}\right].
\label{critical_line}
\end{equation}
Moreover, we have $q_4=8(1+b)^3/27$ under critical conditions for any $b>-1$, ensuring that $\widetilde{\varphi}_i$ in fact reaches
a minimum at the critical points.

\begin{figure}
\flushright
\includegraphics[scale=1.2]{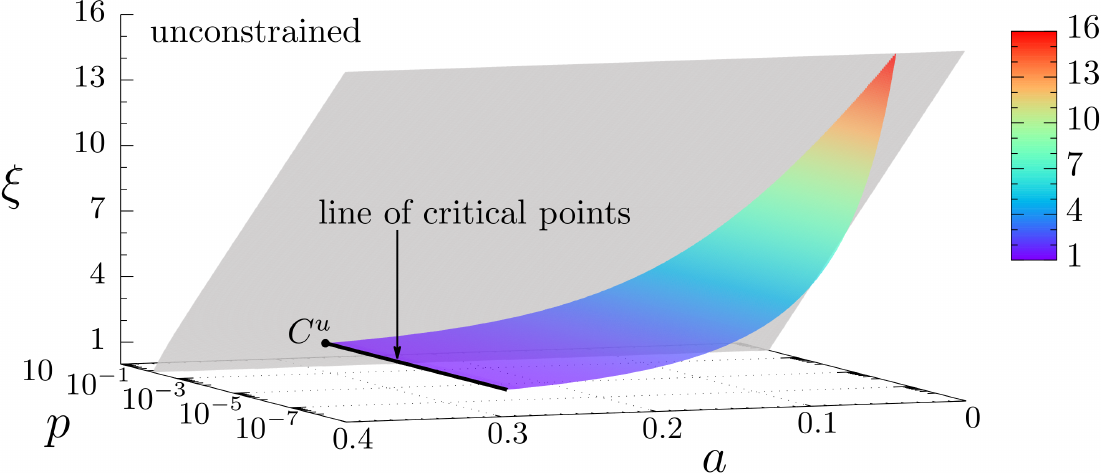}
\caption{Phase diagram $(p,a,\xi)$ in the unconstrained ensemble for any coupling $b<0$. The gray surface indicates the upper bound
$\xi_\mathrm{sup}(p,a)$ below which equilibrium states exist in this ensemble for given $p$ and $a$. The
first-order transition surface in colors separates the phase with high
concentration of particles in
the core (below the surface) from the phase with low concentration (above the surface). This surface ends at a line of critical points
indicated in black, which in turn terminates at the point $C^u=\left(0.2706,\frac{8}{27},\frac{7}{4}-\ln \frac{27}{16}\right)$.}
\label{fig3}
\end{figure}

Considering for a moment the case $b=0$ (the original Thirring model with excluded volume effects), we note that equation
(\ref{critical_line}) can be numerically solved for $b=0$ to give $p^i= 0.2706$. Then the critical point in this case is
characterized by $x_0^i=9/8$, $a^i=8/27$ and $p^i= 0.2706$ and it does not depend on the
reduced total number particles $x$.  In other terms, for $b=0$ the critical line in the phase diagram $(p,a,x)$ is a line parallel
to the $x$ axis. Restoring back the system parameters from the reduced variables, we get the critical values of
these parameters at the isothermal-isobaric critical point for $b=0$:
\begin{eqnarray}
N_0^i &=& \frac{ V_0}{ 3 \sigma},\\
T^i &=& \frac{8 \nu V_0}{ 27 \sigma} ,\\
P^i &=& 0.0238 \frac{\nu V_0}{ \sigma^2}.
\end{eqnarray}

In the case with $b>-1$ and $b\neq0$, the model exhibits a line of critical points given by the relation (\ref{critical_line}) between
$x^i$ and $p^i$, with $x_0^i=9/(8+8b)$ and $a^i=8(1+b)/27$. The line of critical points terminates at $x^i=0$.
In the next section we show a phase diagram $(p,a,x)$ where this line can be appreciated.
In addition, while the critical number of particles in the core $N_0^i$ coincides with
that for $b=0$, the critical temperature now reads 
\begin{equation}
T^i = \frac{8(1+b) \nu V_0}{27 \sigma}.
\end{equation}
Finally, using the above result, the line of critical points (\ref{critical_line}) becomes
\begin{equation}
 N^i ( P^i )= 
\frac{ \sigma P^i }{ 2 b \nu}
+\frac{4(1+b) V_0}{27 b \sigma }\ln\left[ \frac{ 27 \sigma^2 P^i}{ 4 (1+b) \nu V_0} \right]
+\frac{7(1+b) V_0}{27 b \sigma }
\label{critical_line_2}
\end{equation} 
when written in terms of the critical total number of particles $N^i$ and pressure $P^i$.

\begin{figure}
\flushright
\includegraphics[scale=1.2]{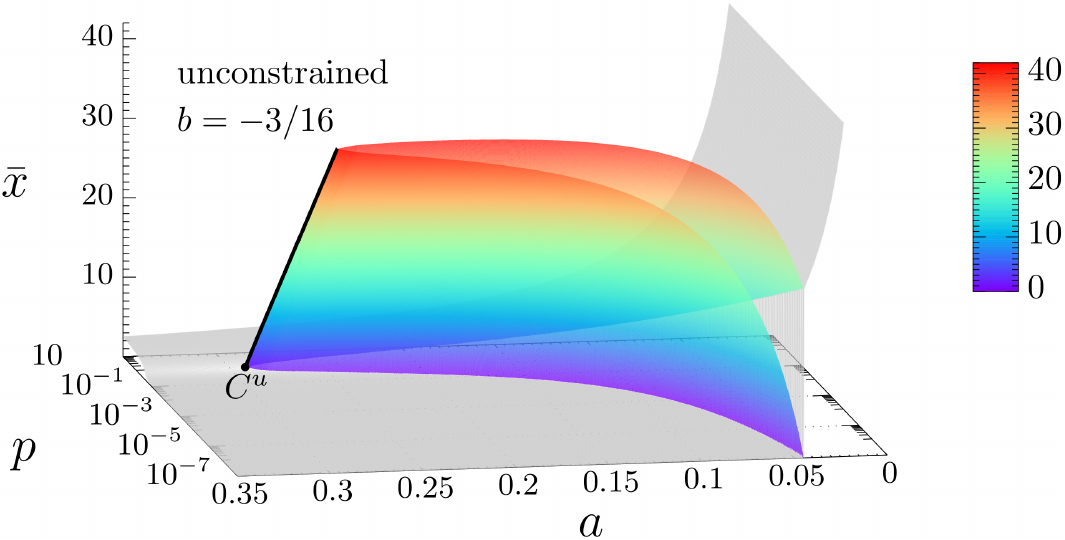}
\caption{Phase diagram $(p,a,\bar{x})$ in the unconstrained ensemble with $b=-3/16$. The gray surface indicates the limiting value for $\bar{x}$ above
which equilibrium states exists in this ensemble for given $p$ and $a$. The colored surfaces represent the discontinuity
in $\bar{x}$ due to the transitions. These surfaces merge at the line of critical points (in black). In this diagram,
$C^u=\left(0.2706,\frac{8}{27},\frac{9}{8}\right)$.}
\label{fig4}
\end{figure}

\section{Phase diagrams}
\label{phase_diag}

In the previous section we obtained the critical points at which first order transitions
terminate. Here we present the full characterization of phase transitions in the system for both the isothermal-isobaric and
unconstrained ensembles. We recall that the latter is equivalent to the grand canonical ensemble for $b<0$, which is the case we
consider here. In this way, phase diagrams in different ensembles can be compared with each other and insight can be gained into
the behaviour of the system when subject to different control parameters. We will show figures for both the full 3D phase
diagram and for some 2D cross sections. The latter can be useful to visualize with more ease some of the properties of the diagrams.
In the full 3D diagrams, the first-order transitions are located on surfaces ending on critical lines, while in the 2D cross sections
they are characterized by lines ending in a critical point. Accordingly, in the following we will refer to first-order transition
surfaces or lines depending on the dimension on the considered case, 3D or 2D phase diagram.

\begin{figure}
\flushright
\includegraphics[scale=0.85]{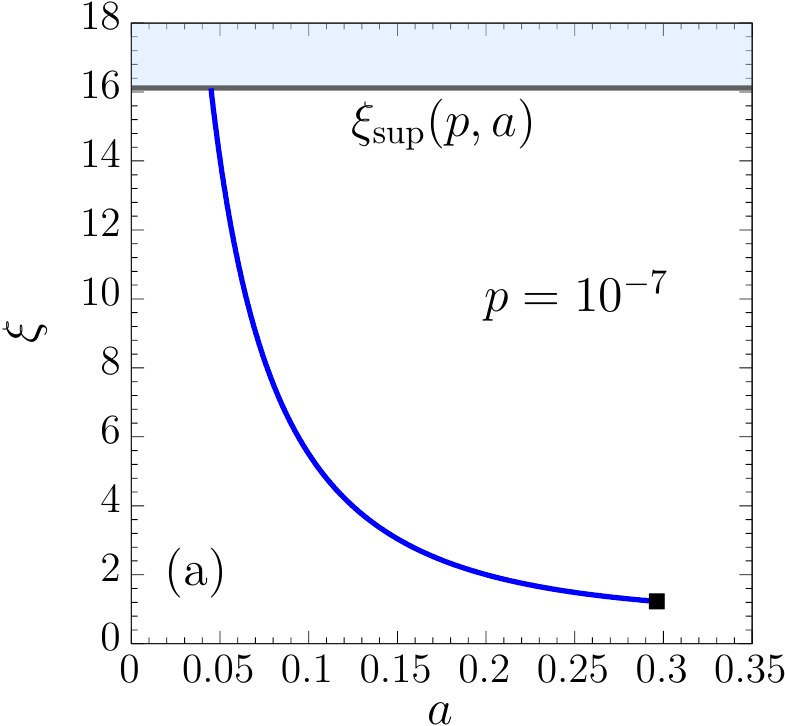}~\includegraphics[scale=0.85]{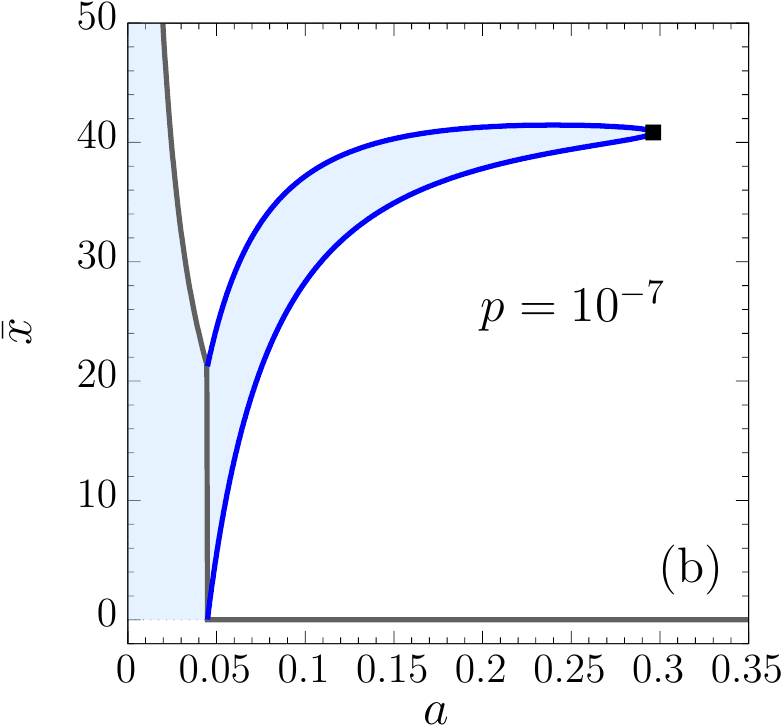}\vspace{3mm}
\includegraphics[scale=0.85]{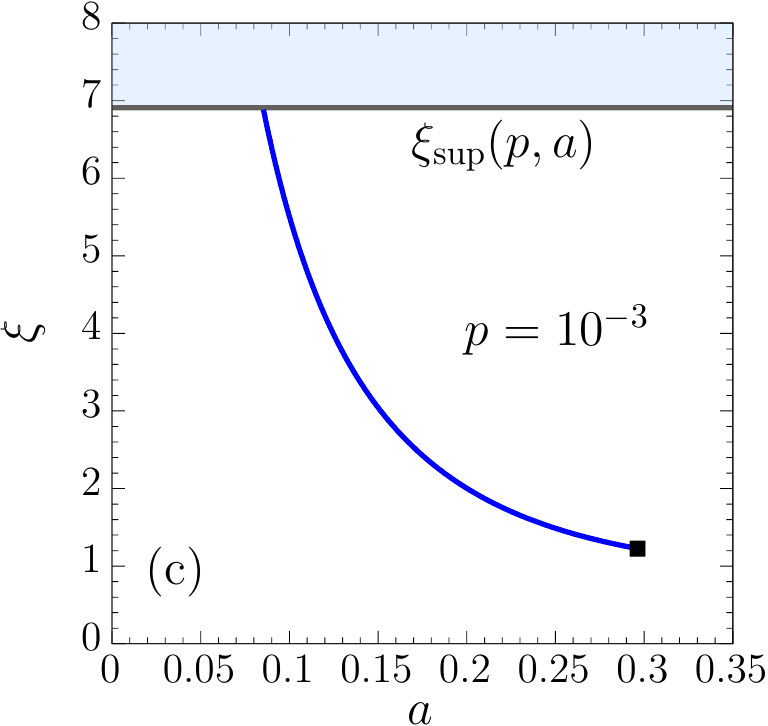}~\includegraphics[scale=0.85]{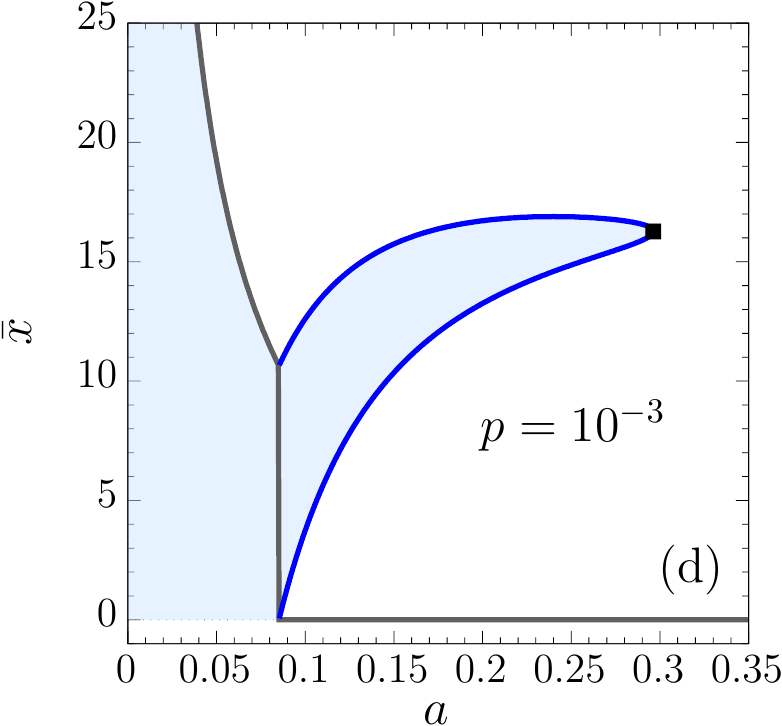}
\caption{Cross sections of the phase diagram in the unconstrained ensemble. In (a) and (c), $\xi_\mathrm{sup}(p,a)$ (gray line)
sets an upper bound for $\xi$. A first-order transition line is shown (in blue) that terminates at a
critical point (black square).
In (b) and (d), we take $b=-3/16$ and the gray lines show the limiting case $\bar{x}=\bar{x}_0$ which corresponds to $\xi$ approaching
the bound $\xi_\mathrm{sup}(p,a)$. The blue lines in these figures correspond to the two branches accounting for the discontinuity
in $\bar{x}$ at the transition. This two branches merge at the critical point represented by a black square. The light-blue area in
the figures indicates a region of non-realizable points in equilibrium configurations. }
\label{fig5}
\end{figure}

While the location of critical points has been determined analytically using Landau's theory, the location
of first order transitions, in general, have to be found numerically. We focus first on the unconstrained ensemble. In this ensemble,
the location of first order transitions in the space of control parameters is defined by points making
equal the two minima of the reduced replica energy (\ref{res_replica_energy}). The minimization of the replica energy does not depend
on the reduced pressure $p$ and the parameter $b$, as can be seen from (\ref{res_replica_energy}). However, for given $p$ and exclusion
parameter $a$, the reduced chemical potential $\xi$ in equilibrium configurations is bounded by the condition (\ref{condition_pressure}),
which can be written as
\begin{equation}
\xi< \xi_\mathrm{sup}(p,a),\qquad \xi_\mathrm{sup}(p,a) = - \ln \left( p \ee^{a p}\right).
\end{equation}
As discussed in Section~\ref{uncon_ens}, this condition ensures that the reduced volume and the number of particles in the outer region
of the system are positive quantities. Thus, $\xi_\mathrm{sup}$ restricts the space of parameters and the phase diagram as well.

In figure~\ref{fig3}, we show the phase diagram in the space of parameters $(p,a,\xi)$ for the unconstrained ensemble. As explained
before, the minimization of the replica energy does not depend on $b$, so that this diagram correspond to any coupling $b<0$. The gray
surface in the figure indicates the bound $\xi_\mathrm{sup}(p,a)$ for $\xi$ below which equilibrium states exists in this ensemble
for given $p$ and $a$. The first-order transition surface in colors separates
the phase with high values of $\bar{x}_0$ (below the surface) from the phase with low values of $\bar{x}_0$ (above the surface). As
already noted, this surface
terminates at a line of critical points indicated in black in the figure, and the point
$C^u=\left(0.2706,\frac{8}{27},\frac{7}{4}-\ln \frac{27}{16}\right)$ characterizes the endpoint of the line. Moreover, the jump
in $\bar{x}_0$ at the transitions induces a jump in the total reduced number of particles $\bar{x}$. In figure~\ref{fig4}, we map
the previous phase diagram $(p,a,\xi)$ into a phase diagram in the space of parameters $(p,a,\bar{x})$, where the discontinuity
in $\bar{x}$ is indicated by the surfaces in color. In this case the location of the surfaces of first order transitions
and that of the critical points does depend on $b$, since, as can be seen from equation (\ref{eq_3}), the value of $\bar{x}_1$ depends
on $b$. In the example of this figure, we set $b=-3/16$. The bound $\xi_\mathrm{sup}(p,a)$ is mapped into the gray surface shown in
the figure; when $\xi$ approaches $\xi_\mathrm{sup}(p,a)$, $\bar{x}$ approaches $\bar{x}_0$ for the given values of $p$ and $a$, so
that the gray surface represents the limiting situation $\bar{x}=\bar{x}_0$. In the phase diagram $(p,a,\bar{x})$, the endpoint $C^u$
is mapped into $C^u=\left(0.2706,\frac{8}{27},\frac{9}{8}\right)$.

\begin{figure}
\flushright
\includegraphics[scale=1.2]{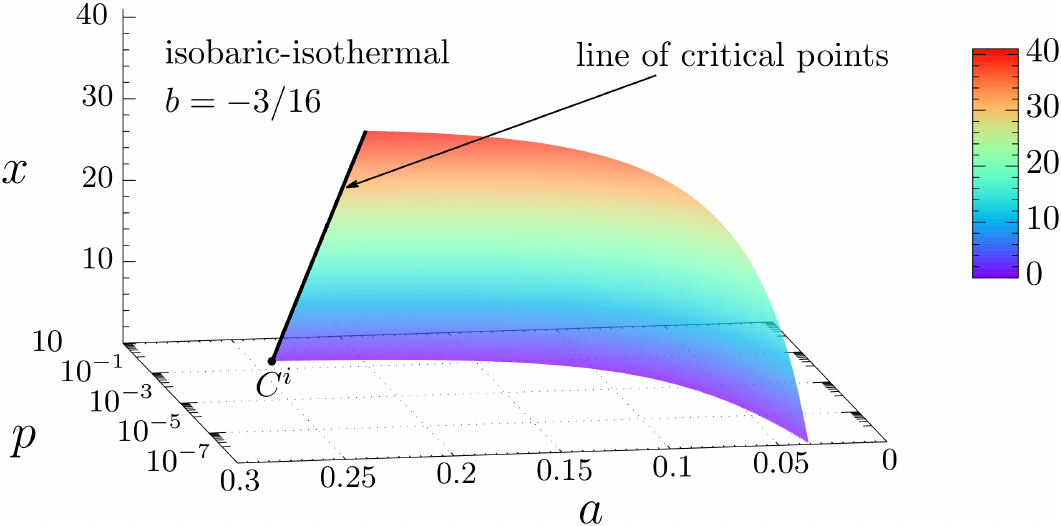}
\caption{Phase diagram $(p,a,x)$ in the isothermal-isobaric ensemble with coupling $b=-3/16$. The first-order transition surface in colors separates the phase with high concentration of particles in the core (above
the surface) from the phase with low concentration (below the surface). This surface ends at a line of critical points indicated in
black, and the line finishes at the point $C^i=\left(\frac{0.2706}{1+b},\frac{8(1+b)}{27},0\right)$.}
\label{fig6}
\end{figure}

To give more details about this phase diagram, in figure~\ref{fig5} we show a cross section on the planes $(a,\xi)$ and $(a,\bar{x})$
for some fixed values of the pressure ($p=10^{-3}$ and $p=10^{-7}$). In the representation of the diagram with parameters $(a,\xi)$,
figures~\ref{fig5}(a) and \ref{fig5}(c), the bound $\xi_\mathrm{sup}(p,a)$ (gray line) sets an upper bound for $\xi$ and thus,
the region above this bound contains forbidden values for the reduced chemical potential in equilibrium configurations. A first-order transition
line (in blue) is shown in these figures which terminates at a critical point represented by a black square. In the cross section
of the phase diagram on the plane $(a,\bar{x})$, figures~\ref{fig5}(b) and \ref{fig5}(d), the first order transition
line is mapped into the two branches (blue lines) accounting for the discontinuity of $\bar{x}$ at the transition, which merge and end
at the critical point (black square). Moreover, the bound $\xi_\mathrm{sup}(p,a)$ corresponds to the gray line in these figures, which
defines a region with forbidden points in the plane $(a,\bar{x})$. The region in between the two branches contains non-realizable points
as well.

\begin{figure}
\flushright
\includegraphics[scale=0.85]{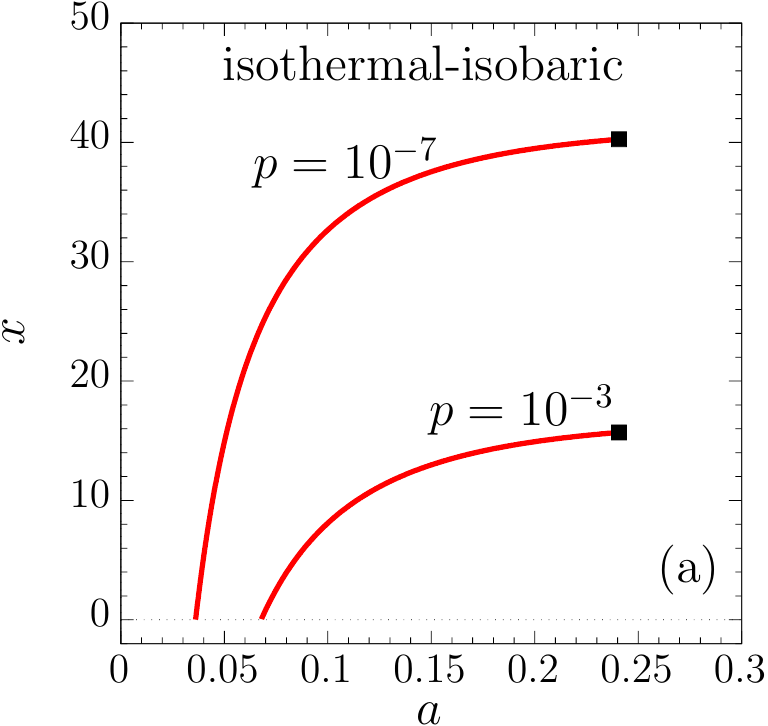}~\includegraphics[scale=0.85]{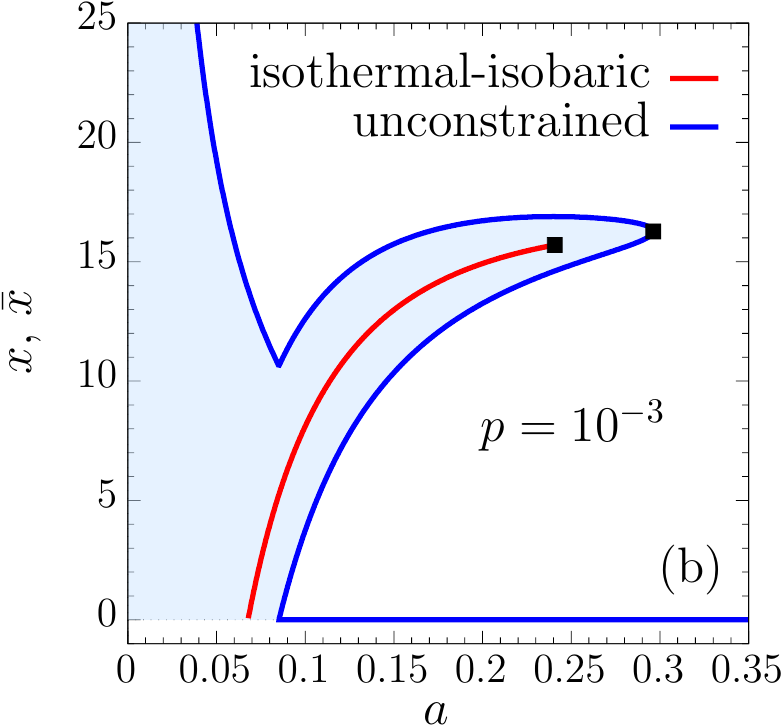}
\caption{Cross sections of the phase diagram. Here we take $b=-3/16$. In (a), two lines of first-order transitions are shown in the isothermal-isobaric
ensemble on the plane $(a,x)$. In (b), a line of first-order transitions (in red) in the isothermal-isobaric ensemble is compared with a cross section
of the unconstrained phase diagram, in which the blue line is the union of the branches accounting for the discontinuity in $\bar{x}$
and the bound defining permitted values of $\bar{x}$, see figure~\ref{fig5}(d). The light-blue area specifies non-realizable points
in the unconstrained ensemble. In both figures, black squares denote critical points.}
\label{fig7}
\end{figure}

We now focus on the isothermal-isobaric ensemble. The location of first-order transitions in this case is
defined by points making equal the two minima of the reduced Gibbs free energy (\ref{varphi_I_2}).
In figure~\ref{fig6}, the phase diagram $(p,a,x)$ is shown for this ensemble with coupling $b=-3/16$. The
first-order transition surface in colors separates the phase with high
concentration of particles in
the core (above the surface) from the phase with low concentration (below the surface). The transition surface ends at a line of critical
points indicated in black. This line finishes when the reduced number of particles vanishes, $x=0$. Imposing this condition in the line
equation (\ref{critical_line}) yields
\begin{equation}
\ln\left( a^i p^i \right) + a^i p^i + \frac{7}{4} + \ln 2=0,
\label{critical_line_3}
\end{equation}
where $a^i=8(1+b)/27$, and whose (numerical) solution is $p^i=0.2706/(1+b)$. Thus, the endpoint of the line of critical points in the
isothermal-isobaric ensemble is $C^i=\left(\frac{0.2706}{1+b},\frac{8(1+b)}{27},0\right)$, which is shown in figure~\ref{fig6}
for $b=-3/16$. In figure~\ref{fig7}(a), two cross sections of the phase diagram are represented on the plane $(a,x)$ corresponding
to $p=10^{-3}$ and $p=10^{-7}$. There, the first order transition lines terminate at critical points denoted with
black squares. 

Finally, in figure~\ref{fig7}(b) we compare a cross section of the isothermal-isobaric phase diagram with a cross section of the phase
diagram in the unconstrained ensemble, for $b=-3/16$ and $p=10^{-3}$. The cross sections correspond to points $(a,x)$ in the
isothermal-isobaric ensemble and $(a,\bar{x})$ in the unconstrained case. In the figure, the red line represents a
first-order transition line in the isothermal-isobaric ensemble, while the blue line is the union of the branches accounting for the
discontinuity in $\bar{x}$ and
the bound defining permitted values of $\bar{x}$ in the unconstrained ensemble, that is, the union of the blue and gray lines in
figure~\ref{fig5}(d). The light-blue area shown in the figure specifies non-realizable points in the unconstrained ensemble, and
the black squares denote the critical points in the corresponding ensembles. We observe that the critical points do not coincide
because the two ensembles are not equivalent. Moreover, the isothermal-isobaric first-order transition line is
well deep in the region of the unconstrained phase diagram where there are no realizable equilibrium states in this ensemble, which
highlights the inequivalence of the ensembles.

\section{Discussion and conclusions}
\label{discuss}

In this paper we have studied the phase diagram of the modified Thirring model with particles of finite size in the unconstrained
ensemble. We have shown that the model exhibits phase transitions in this ensemble. The comparison with the isothermal-isobaric ensemble
has revealed that the two ensembles are inequivalent. Also in the latter ensemble the system has phase transitions, but they manifest
themselves in a different way and at different values of the thermodynamic variables. In particular, the surfaces in the 3D thermodynamic
phase diagram in the two ensembles are distinct from each other, and so are the lines of critical points. Looking at figure \ref{fig2} one recognizes
similar features to those encountered in the study of ensemble inequivalence between microcanonical and canonical ensembles (the case
where most of the studies of ensemble inequivalence have considered): the ``less constrained'' (or ``higher'') ensemble, in our case the
unconstrained ensemble, is characterized by a first-order phase transition with a Maxwell construction (the horizontal section of the
red line in the figure), while the first-order transition in the ``more constrained'' (or ``lower'') ensemble, the isothermal-isobaric
one, presents a jump in the dependent (i.e., not control) thermodynamic variable $\xi$ together with a range with anomalous sign
of the derivative. On the other hand, we have found that the grand canonical ensemble is equivalent to the unconstrained ensemble.

We emphasize that phase transitions in this model are ``not-symmetry-breaking'' transitions, similarly to what occurs in a
liquid-gas transition, and differently from what is found in a magnetization transition \cite{Bouchet_2005,Cohen_2012}: in the latter
the transition is characterized by the passage from a unmagnetized state, that enjoys an up-down symmetry, to a magnetized state without
this symmetry; in the former no such symmetry is broken at the transition, which is caracterized by a jump in the density. In our model,
at the transition we have a similar jump in density, when the reduced number of particles $x_0$ in the core region change abruptly by
varying, e.g., $\xi$ in the unconstrained ensemble or $p$ in the isothermal-isobaric ensemble. In the latter ensemble, where the total
number of particles is a fixed control parameter, the jump in $x_0$ is accompanied by a corresponding jump in $x_1$, the reduced number
of particles in the outer region. In a 3D phase diagram, the symmetry-breaking transitions are characterized by surfaces of continuous
(or second order) transitions, contrary to what is found in systems with transitions that do not break a symmetry, where the the second-order transitions lie on lines of critical points. In a 2D section of the phase diagram, e.g., a $(a,\xi)$ diagram in our reduced
variables, we thus find a line of first-order transitions terminating at a critical point, similarly to a liquid-gas transition.

Another important point in the passage from a ``lower'' to a ``higher'' ensemble is that generally the number of accessible states of
the system decreases \cite{Ellis_2000}. This is related to the Maxwell construction at the first-order phase transitions in
the ``higher'' ensemble\footnote{We recall that in nonadditive systems the physically realized states in the Maxwell construction are
only those at the two extremes of the corresponding segment, while states in the internal points are not realized. This is due to the
fact that phase separation in such systems does not occur, at variance with additive systems, where the states corresponding to the
internal points can be realized by the separation of the system in two phases.}. This decrease in the number of accessible states can
cause also the presence of ranges in the parameters of the system where equilibrium states are not found at all.
This is the case that occurs for our system in the unconstrained ensemble, that has equilibrium states only for $b<0$, while the
isothermal-isobaric ensemble and the grand canonical ensemble have equilibrium states also for for $b>0$ (the equivalence between
unconstrained and grand canonical ensembles considers of course only the case $b<0$, where there are equilibrium states in both
ensembles).

In a previous paper \cite{Latella_2017} we have studied the system with point-like particles, i.e., with $\sigma=0$ (so that also the
reduced variable $a$ identically vanishes). In that case we had found that the unconstrained
ensemble has equilibrium states only for $b<0$, similarly to what has been obtained now, but contrary to the present results, there are no phase transitions. Also in that case
we had found equivalence with the grand canonical ensemble. Therefore, the introduction of a finite size has changed dramatically the
behaviour of the system. The finite size is equivalent to the presence of a hard core in the interaction between the particles. The main
consequence is un upper bound, equal to $V/\sigma$, in the possible number of particles in a given volume $V$. It is known that
unusual thermodynamic behavior can occur when the interaction is not stable \cite{ruelle_1969}, where an interaction is defined to be
stable when in all possible configurations of $N$ particles the total potential energy is lower bounded by $-BN$, where $B$ is an $N$-independent positive constant. Two possible reasons that can prevent stability are: an attractive two-body interaction that is singular
at the origin (or even a not sufficiently repulsive two-body interaction at the origin \cite{ruelle_1969}), and an attractive long-range
interaction. The gravitational interaction, but also the original Thirring model and the modified Thirring model studied
in \cite{Latella_2017}, have both features. On the other hand, in the modified Thirring model with finite size
particles, as considered in this work, only the second feature is present. The main physical consequence of this fact is the upper
bound $V_0/\sigma$ in the number of particles in the core, and that trying to accommodate a number of particles close to this bound produces
a strong decrease of the entropy, and hence a strong increase of the free energies as well. This becomes clear by looking at, e.g., equations (\ref{res_replica_energy}), (\ref{varphi_I_2}) and (\ref{reduced_grand_potential}),
where we see that the free energies diverge to $+\infty$ when $x_0$ approaches the upper bound $1/a$. Because of this, it is
possible for the free energy to have a second minimum for large $N_0$, i.e., $x_0$ close to $1/a$, besides that for small $N_0$.

Furthermore, if $\sigma$ vanishes, the minimum at large $N_0$ takes place at $N_0 \to \infty$. As a matter of fact, when $\sigma=0$, the global minimum
(considering the dependence on $N_0$) of the free energies in the unconstrained and grand canonical ensembles happens at
$N_0 \to \infty$, where these free energies tend to $-\infty$. We stress that the number of particles is not fixed in these ensembles.
In other words, these global minima occur for a collapsed state, although
this collapse is not of the same nature of that in self-gravitating systems; in the latter all particles collapse to the same point,
while in this system a diverging number of particles would stay in the core $V_0$. In this respect, the equilibrium states found
in \cite{Latella_2017} for the unconstrained and grand canonical ensembles have to be regarded, from a strict thermodynamical point of view, as metastable states. However, we have
to consider that the lifetime of metastable states in self-gravitating systems, for which the global minimum of the free energy
is represented by a completely collapsed state, diverges exponentially with the number of particles \cite{Chavanis_2006}.
So, for large systems with point-like particles ($\sigma=0$), such metastable states can be considered physically as equlibrium states.
The hard core introduced by a finite $\sigma$, as considered here, avoids metastability and the associated collapsed states with a diverging number of particles. Thus, our model represents a nonadditive, macroscopic system where both equilibrium states and phase transitions are present under completely open conditions.

In addition, as mentioned in Section~\ref{local_eos}, we note that a study of particles with hard core interacting through gravitational forces was performed long time ago in the framework
of the canonical ensemble \cite{Aronson_1972}. In that paper, the authors showed that the presence of both a fixed volume enclosing
the system and a hard core lead to the existence of equilibrium thermodynamic states and of phase transitions between states
with approximately uniform densities and states with a large density of particles in the central region of the volume. These
results bear a clearly similarity to those we have obtained in the unconstrained, isothermal-isobaric and grand canonical ensembles.

An interesting problem related to the present work would be the study of the relaxation times of the system towards equilibrium states, for instance, by Monte Carlo
simulations. One could investigate if also these times are ensemble dependent. Since we are not considering isolated systems (treated
by the microcanonical ensemble), where long-lived quasistationary states are very often found with long-range interactions, one might
argue that the relaxation times to equilibrium are similar, independently of allowing only fluctuations of $E$ among the constraint variables, or also of $V$ and/or $N$; but this could be confirmed only by a dedicated study.

\section*{Acknowledgments}
AC acknowledges financial support from INFN (Istituto Nazionale di Fisica Nucleare) through the projects DYNSYSMATH and ENESMA.

\end{document}